\documentclass[runningheads]{llncs}
\usepackage[T1]{fontenc}
\usepackage{graphicx}
\usepackage{multirow}%
\usepackage{amsmath,amssymb,amsfonts}%
\usepackage{mathrsfs}%
\usepackage[title]{appendix}%
\usepackage{xcolor}%
\usepackage{textcomp}%
\usepackage{manyfoot}%
\usepackage{booktabs}%
\usepackage{algorithm}%
\usepackage{algorithmicx}%
\usepackage{algpseudocode}%
\usepackage{listings}%
\usepackage{tikz}
\usetikzlibrary{positioning,arrows.meta,shadows}
\usepackage{makecell}
\usepackage{hyphenat}
\usepackage{hyperref} 
\usepackage[english]{babel}

\begin{document}
\title{RAGEAR: Retrieval-Augmented Graph-Enhanced Academic Recommender}
%
%
\author{
Francesco Granata\inst{1}\orcidID{0009-0001-3023-3265} \and
Lorenzo Lamazzi\inst{2}\orcidID{0009-0004-9808-2936} \and
Misael Mongiovì\inst{1,2}\orcidID{0000-0003-0528-5490} \and
Francesco Poggi\inst{2}\orcidID{0000-0001-6577-5606} \and
Valeria Secchini\inst{2}\orcidID{0000-0002-8189-7459} 
}
\authorrunning{ F. Granata et al.}
%
\institute{ 
Department of Mathematics and Computer Science, University of Catania, Italy\\ \and
Institute for Cognitive Science and Technology, National Research Council, Italy 
}
\maketitle              
\begin{abstract}

\ We \ present\  RAGEAR \ (Retrieval-Augmented \ Graph-En- hanced Academic Recommender), a neurosymbolic recommender system for academic course recommendation. RAGEAR combines dense retrieval over full lecture transcripts with a symbolic Knowledge Graph modelling courses, lessons, transcript chunks, credits, study plans, and curricular information. The Knowledge Graph supports symbolic filtering and contextualisation based on structured constraints, such as credits, academic disciplines, study plans, and prerequisites. Unlike metadata-based approaches, it exploits fine-grained instructional content by retrieving transcript chunks semantically aligned with a student's query.
The main contribution is a graph-aware aggregation function that propagates chunk-level evidence to course-level recommendations. The score combines three factors: the share of retrieved similarity associated with a course, the rank-based strength of its relevant chunks, and the distribution of evidence across lessons.
We evaluate RAGEAR on 152 student-like queries through a human evaluation sample and a large-scale LLM-based relevance assessment. Results show that lecture transcripts improve over metadata-only retrieval, and that RAGEAR further improves ranking quality over a transcript-based normalized SumP baseline, especially for top-ranked recommendations.

\keywords{Course Recommendation \and Knowledge Graph \and Ontology Engineering \and Neurosymbolic Recommender Systems \and Dense Retrieval}
\end{abstract}
\section{Introduction}\label{sec1}

Recommender systems are increasingly used to help users navigate large and heterogeneous information spaces. In higher education, they can support students in selecting learning resources, elective courses, degree programmes, or coherent study paths. This is particularly useful when students must choose among many courses described through heterogeneous and often high-level information.

Academic course recommendation is challenging because students usually express their interests through natural-language queries that do not necessarily match official course descriptions. For example, a student may ask for courses related to practical data analysis, secure software development, scalable online services, or the legal implications of data processing, without using the same terminology found in course titles, abstracts, or syllabus metadata. As a result, recommenders based only on course-level metadata may capture broad thematic similarity, but fail to identify fine-grained conceptual evidence that appears only in the actual teaching material.

This limitation is especially relevant in online universities, where complete video lectures and transcripts are available. Lecture transcripts provide a much richer description of course content than metadata alone: they include examples, technical terminology, explanations, and topic developments across lessons. However, exploiting this material requires moving from coarse course-level representations to fine-grained retrieval over lecture chunks, and then aggregating the retrieved evidence into meaningful course-level rankings.

We present RAGEAR, a Retrieval-Augmented Graph-Enhanced Academic Recommender for academic course recommendation. RAGEAR follows a hybrid neural and symbolic design. The neural component performs dense retrieval over transcript chunks to identify lecture fragments semantically aligned with a student's query. The symbolic component is an ontology-based Knowledge Graph modelling courses, lessons, transcripts, chunks, credits, study plans, academic disciplines, prerequisites, and curricular information. This symbolic layer connects fine-grained lecture evidence with the academic structure of the curriculum.

The Knowledge Graph also supports symbolic filtering and contextualisation of recommendations according to structured constraints or user desiderata, such as credits, academic disciplines, study plans, prerequisites, and course-level requirements. Since such filtering is deterministic once the relevant information is represented in the graph, our empirical evaluation focuses on the ranking component of the system: whether lecture-transcript retrieval and the proposed aggregation function improve recommendation quality over metadata-only retrieval and simpler transcript-based aggregation.

The main methodological contribution of RAGEAR is a graph-aware aggregation function that propagates chunk-level retrieval scores to course-level recommendations. The score combines the share of retrieved similarity associated with a course, the rank-based strength of its retrieved chunks, and the distribution of relevant evidence across lessons. This favours courses whose relevance is supported by multiple highly ranked pieces of evidence distributed across the course structure.

We evaluate RAGEAR by comparing three methods: a metadata-only dense retrieval baseline, a transcript-based normalized SumP aggregation baseline, and the full RAGEAR aggregation. The evaluation combines a human assessment sample, used to measure agreement with an LLM-based relevance judge, and a large-scale LLM-based evaluation on 152 student-like queries. Results show that lecture transcripts improve over metadata-only retrieval, and that the full RAGEAR aggregation further improves ranking quality.

In summary, this paper contributes: (i) a hybrid neural and symbolic architecture for academic course recommendation; (ii) a graph-aware aggregation function for propagating transcript-level evidence to course-level rankings; and (iii) an empirical evaluation against metadata-only retrieval and normalized SumP aggregation.

The remainder of this article is organized as follows. Section \ref{sec:related} reviews related work on course recommendation, educational retrieval, and knowledge-graph–based systems. Section \ref{sec:dataset} describes the dataset and transcript processing. Section \ref{sec:methodology} presents the RAGEAR methodology. Section \ref{sec:evaluation} reports the evaluation protocol and results. Finally, Section \ref{sec:conclusions} concludes the paper and outlines directions for future work.


\section{Related Work}\label{sec:related}

Research on course recommendation has developed along multiple methodological directions, including performance prediction, preference modelling, and multi-criteria decision support. Several systems aim to support students in selecting elective or optional courses based on structured features. Malhotra et al.~\cite{malhotra2022} cluster learners according to interest areas and use matrix factorisation within each cluster to predict expected performance in elective courses. Esteban et al.~\cite{esteban2020} propose a hybrid multi-criteria recommender that integrates collaborative and content-based filtering, optimising the contribution of academic indicators, personal interests, and course workload through a genetic algorithm. Fernández-García et al.~\cite{fernandez2020} design a content-based decision support tool to assist subject enrolment, drawing on course descriptors and indicators of academic risk. Sankhe et al.~\cite{sankhe2020} follow a skills-oriented approach, recommending courses by clustering students and alumni using fuzzy methods and comparing their learning trajectories. These systems rely on course-level metadata, grades, skills, or historical performance, and do not examine the instructional content of courses in fine detail. For a broader discussion of course recommendation methodologies, see~\cite{algarni2023systematic}.

A second line of work introduces ontologies and knowledge-based representations to enrich course recommendation with explicit semantic structure. Obeid et al.~\cite{obeid2018} develop an ontology modelling programmes, prerequisites, and learner characteristics, enabling semantic reasoning to filter feasible courses. Ibrahim et al.~\cite{ibrahim2019} extend this idea through OPCR, a hybrid ontology-based framework combining concept-level similarity with collaborative and content-based filtering. Li et al.~\cite{li2024} introduce CourseKG, a large-scale educational Knowledge Graph capturing relations between courses, competencies, and learning objectives to enable personalised learning-path recommendations. These systems take advantage of semantic modelling and structured curricular information, but their reasoning remains at the level of metadata.

Beyond course-level recommendation, several areas of educational technology investigate semantic matching between learning resources and user queries. Semantic retrieval has been used to align student questions with relevant textbook sections or learning objects using rules engines~\cite{hucontext}. These approaches demonstrate the usefulness of deep content analysis in educational settings. Hybrid approaches combining knowledge graphs with neural embeddings have also gained traction in personalised education. Nguyen et al.~\cite{thuknow} fuse KG-based structure with embedding-based similarity for personalised recommendation of learning resources, and Davis et al.~\cite{davisedu} propose a system that leverages the affordances of KGs, LLMs, and pedagogical agents for personalized learning. These works highlight the potential of unifying symbolic constraints with semantic similarity.


Compared with previous work, RAGEAR combines two aspects that are usually addressed separately: symbolic modelling of academic structures and fine-grained retrieval over full lecture transcripts. Existing course recommenders typically rely on metadata, grades, skills, or learning objectives, whereas our approach ranks courses according to the semantic alignment between a student's query and the actual instructional content. The Knowledge Graph provides curricular structure and constraint-aware contextualisation, while transcript-level retrieval supplies fine-grained evidence for course recommendation.



\section{Dataset}\label{sec:dataset}

The dataset integrates heterogeneous information describing university courses, lecturers, study plans, credits, Academic Disciplines (AD)\footnote{Academic Disciplines classify the teaching and scientific activities of Italian professors and researchers - complete list available at: \url{https://www.cun.it/uploads/storico/settori_scientifico_disciplinari_english.pdf}.}, teaching material, lessons, and associated video recordings. These data are used to populate the ontology-based Knowledge Graph described in Section~\ref{sec:kg}. The corpus includes courses from two Italian online universities and covers 34 courses belonging to one bachelor's degree programme and two master's degree programmes in Computer Science and Engineering.

The courses include 1165 video lectures, corresponding to approximately 821 hours of recorded instructional content. All videos were transcribed using Whisper Turbo~\cite{radford2023robust}, producing full lecture transcripts and word-level timestamps. This temporal information enables the alignment of textual fragments with their original video intervals.

For retrieval, each transcript was segmented into coherent textual units, referred to as chunks. Sentence segmentation was performed with spaCy~\cite{explosion_spacy_2023}, and chunk length was constrained to remain within a semantically meaningful range. For each chunk, start and end timestamps were computed by aligning its token sequence with the Whisper word-level timing information.

Each chunk is associated with its course, lesson, transcript, textual content, index, and temporal metadata. These elements are represented in the Knowledge Graph through the Course--Lesson--Transcript--Chunk structure. This representation supports dense retrieval over fine-grained instructional content and allows the system to propagate chunk-level retrieval scores to course-level recommendations through the RAGEAR aggregation function.

The resulting dataset is a temporally grounded corpus of university lecture transcripts, enriched with structured academic metadata and integrated into an ontology-based Knowledge Graph.

\section{Methodology}\label{sec:methodology}

RAGEAR is a hybrid neural and symbolic architecture for academic course recommendation. Given a student query expressed in natural language, the system retrieves fine-grained evidence from lecture transcripts and aggregates this evidence into course-level recommendations. The neural component performs dense retrieval over transcript chunks, while the symbolic component is an ontology-based Knowledge Graph modelling the academic structure of the learning environment, including students, study plans, courses, lessons, transcripts, chunks, credits, academic disciplines, prerequisites, and curricular information.

The overall workflow involves four main components. First, the user interface collects the student's free-text query and displays the ranked course recommendations. Second, the Knowledge Graph provides the academic context needed to identify and organise candidate courses, their lessons, transcripts, chunks, and structured constraints. Third, the dense retrieval module computes semantic similarity between the query and the transcript chunks associated with the candidate courses. Finally, the recommendation algorithm propagates chunk-level similarity scores to the course level using the RAGEAR aggregation function described in Section~\ref{sec:course-score}.

Given a query $q$, the system can use the Knowledge Graph to restrict or contextualise the candidate set according to structured academic information, such as the student's study plan, academic discipline, credits, prerequisites, or other course-level requirements. The transcript chunks associated with the resulting candidate courses are then passed to the dense retrieval module, which returns the top-ranked chunks according to their semantic similarity with the query.

The retrieved chunk-level scores are subsequently aggregated along the Course--Lesson--Chunk structure represented in the Knowledge Graph. This aggregation favours courses that accumulate substantial retrieval evidence, contain highly ranked relevant chunks, and distribute such evidence across multiple lessons. The system finally returns the top recommended courses together with course metadata and supporting transcript evidence, enabling the student to inspect why a course has been recommended. Figure~\ref{fig:architecture-overview} summarises the workflow of the system.

\usetikzlibrary{positioning,arrows.meta}

\begin{figure}[h]
\centering

\begin{tikzpicture}[
    component/.style={
        rectangle,
        rounded corners,
        draw=black,
        fill=gray!10,
        thick,
        text width=3.5cm,
        align=center,
        minimum height=14mm
    },
    algorithm/.style={
        rectangle,
        rounded corners,
        draw=black,
        fill=gray!20,
        thick,
        text width=3.5cm,
        align=center,
        minimum height=14mm
    },
    arrow/.style={-{Stealth}, thick},
    node distance=14mm
]

\node[component] (ui) 
    {component\\ \textbf{User Interface}\\
    \small(1) Query input\\
    \small(11) Display of recommendations};

\node[component, right=4.6cm of ui] (kg) 
    {component\\ \textbf{Knowledge Graph}\\
    \small(3) Course selection\\
    \small(9) Retrieval of course information};

\node[component, below=1.2cm of kg] (rag) 
    {component\\ \textbf{Dense Retrieval Module}\\
    \small(5) Similarity evaluation\\
    \small(dense score)};

\node[algorithm, left=4.6cm of rag] (rec) 
    {algorithm\\ \textbf{Recommendation Algorithm}\\
    \small(7) Computation of RAGEAR Score};

\draw[arrow] (ui) -- 
    node[midway, above, yshift=1mm] {\small (2) Retrieval of allowed courses} 
    (kg);

\draw[arrow] (kg) to[bend right=18] 
    node[midway, below, yshift=5mm] {\small (10) Course metadata and recommendations} 
    (ui);

\draw[arrow] (kg) -- 
    node[midway, right, xshift=-2.92cm] {\small (4) Candidate transcript chunks} 
    (rag);

\draw[arrow] (rag) -- 
    node[midway, above, sloped, yshift=1mm] {\small (6) Top-$k$ chunk similarities} 
    (rec);

\draw[arrow] (rec) -- 
    node[midway, above, sloped, yshift=1mm] {\small (8) Recommended courses} 
    (kg);

\end{tikzpicture}

\caption{High-level architecture integrating User Interface, Knowledge Graph, Dense Retrieval Module, and Recommendation Algorithm.}
\label{fig:architecture-overview}
\end{figure}
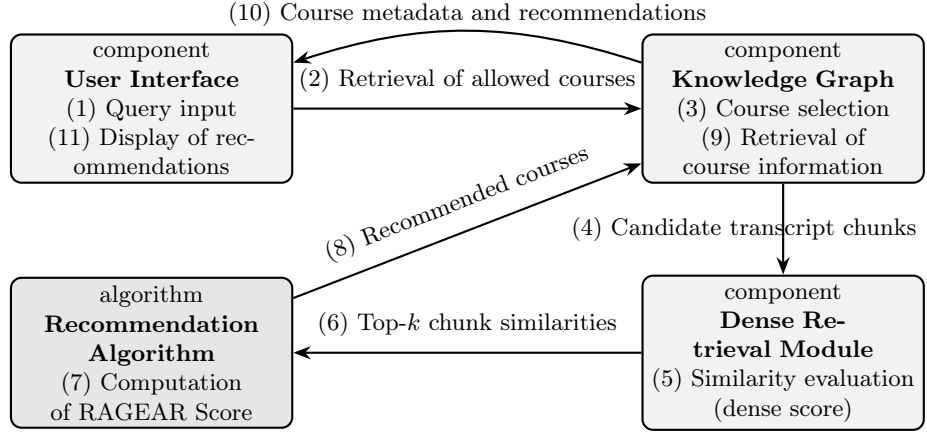



\subsection{Dense Retrieval over Lecture Transcripts}
\label{sec:dense-retrieval}

The retrieval component of RAGEAR identifies the portions of lecture content that are semantically aligned with a student's query. Unlike metadata-based retrieval, which compares the query only with course-level descriptions, this component operates at the level of transcript chunks. This allows the system to capture fine-grained conceptual evidence that may appear in specific lectures even when it is not mentioned in the course title, abstract, or syllabus metadata.

Each lecture transcript is segmented into semantically coherent chunks, as described in Section~3. Each chunk is associated with its course and lesson through the Knowledge Graph, preserving the hierarchical structure Course--Lesson--Chunk. During preprocessing, all chunks are encoded into dense vector representations using the \texttt{multilingual-e5-large} \cite{wang2024multilingual} sentence-transformer model. At query time, the student's natural-language query is encoded with the same model, and semantic similarity is computed between the query embedding and the embeddings of candidate transcript chunks.

Given a query $q$, the dense retrieval module returns the top-$k$ most similar chunks, each associated with a similarity score $S_{cq}$, where $c$ denotes a chunk. In our implementation, we use the top-200 retrieved chunks as the evidence set for course-level recommendation. Chunks outside this set are assigned a score of zero for the purpose of aggregation.

The output of the dense retrieval module is therefore a ranked list of transcript chunks, not a ranked list of courses. Course-level recommendation is obtained in the next step by propagating chunk-level similarity scores through the Course--Lesson--Chunk structure represented in the Knowledge Graph. This aggregation step is described in Section~\ref{sec:course-score}.

\subsection{Ontology Design and Knowledge Graph Structure}
\label{sec:kg}

The symbolic component of RAGEAR is an ontology-based Knowledge Graph (KG) that provides a structured representation of the academic domain and of the instructional material used by the recommender. The KG has two main functions. First, it models the academic context in which recommendations are produced, including students, study plans, courses, credits, academic disciplines, prerequisites, instructors, lessons, transcripts, and transcript chunks. Second, it represents the structural relations needed to propagate retrieval evidence from fine-grained lecture chunks to course-level recommendations.

At the academic level, the ontology models the relation between a student and their study plan. A \texttt{Student} is connected to a \texttt{StudyPlan} through the property \texttt{:hasStudyPlan}, and each study plan is linked to the courses it contains through \texttt{:containsCourse}. Courses are described by structured attributes such as title, credits, academic discipline, prerequisites, and instructors. 
These instances are populated from the university platform data through deterministic mappings from the available course catalogue, study-plan records, lecture metadata, and transcript-processing outputs.
This representation enables symbolic filtering and contextualisation of recommendations according to structured constraints or user desiderata, such as the number of credits, the academic discipline, study-plan compatibility, or prerequisite satisfaction.

\begin{figure}[h!]
    \centering
    \includegraphics[width=0.85\linewidth]{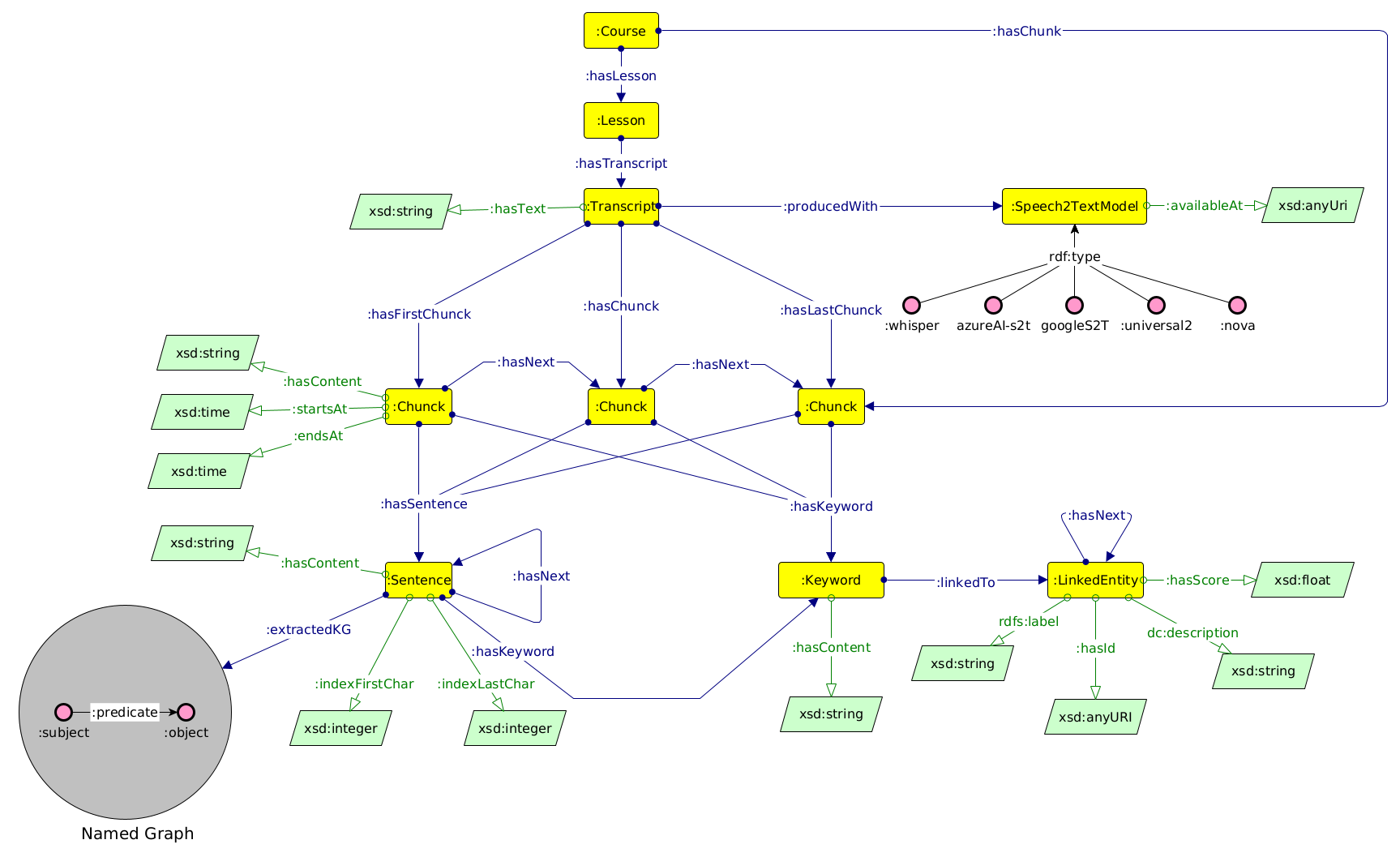}
    \caption{Graffoo diagram depicting the ontology module with information about content, courses and lectures}
    \label{fig:kg_contents}
\end{figure}

At the instructional-content level, the ontology represents courses as structured collections of lessons and transcript chunks. As showed in the Graffoo~\cite{Falco2014320} diagram in Fig.\ref{fig:kg_contents}, a \texttt{Course} is linked to its instructional units and transcripts, and each transcript is decomposed into \texttt{Chunk} instances. Chunks are associated with textual content and temporal metadata, such as their start and end positions in the corresponding lecture video. This modelling preserves the hierarchical structure $\texttt{Course} \rightarrow \texttt{Lesson} \rightarrow \texttt{Chunk}$,
which is central to the RAGEAR aggregation function described in Section~\ref{sec:course-score}. The recommendation algorithm uses this structure to aggregate dense retrieval scores from chunks to courses, while also considering how relevant evidence is distributed across the lessons of a course.

The KG therefore acts as the symbolic backbone of the recommender. It allows the system to move from unstructured transcript fragments to structured academic objects, and to associate each retrieved chunk with its lesson, course, and curricular context. This is essential for producing recommendations that are not only semantically aligned with the student's query, but also interpretable in terms of the academic structure of the curriculum.

The ontology was designed following the eXtreme Design (XD) methodology~\cite{Blomqvist2016}, which supports iterative and competency-question-driven ontology development. Domain specialists defined a set of Competency Questions (CQs) that the KG should answer, such as identifying available courses for a student, retrieving course credits and lecturers, and locating transcript fragments where a concept appears.


The KG also includes an optional sentence-level semantic enrichment layer based on \emph{Named Graphs} \cite{carroll2005named} generated through AMR-to-FRED\cite{Gangemi2026}. This layer is not used as a direct ranking signal in the experiments reported here; it is intended to support future semantic inspection, explanation, and concept-centric exploration.

\subsection{Course Recommendation Score}
\label{sec:course-score}

The dense retrieval module returns a ranked list of transcript chunks for a query
$q$, but the final output of the system must be a ranked list of courses. The
purpose of the RAGEAR scoring function is therefore to propagate chunk-level
similarity evidence to the course level by exploiting the Course--Lesson--Chunk
structure represented in the Knowledge Graph.

Let $R_q$ be the set of the top-200 chunks retrieved for query $q$, and let
$S_{cq}$ denote the dense similarity score between chunk $c$ and query $q$.
Chunks not included in $R_q$ are assigned a score of zero. For each course $C$,
RAGEAR computes a course-level recommendation score as the product of three
components:

\[
RS(C,q) = GE(C,q) \cdot RE(C,q) \cdot LC(C,q).
\]

where $RS$ denotes the course-level recommendation score, $GE$ the global evidence component, $RE$ the ranked evidence component, and $LC$ the lesson coverage component.

The first component, $GlobalEvidence(C,q)$, measures the share of the total
retrieved similarity assigned to chunks belonging to course $C$:

\[
GlobalEvidence(C,q) =
\frac{\sum_{c \in C \cap R_q} S_{cq}}
{\sum_{c \in R_q} S_{cq}}.
\]

This term corresponds to a normalized SumP aggregation over retrieved transcript
chunks. It captures how much of the total retrieval evidence for query $q$ is
associated with course $C$. In the evaluation, this component alone is used as
the \textit{Transcript Normalized SumP} baseline.

The second component, $RankedEvidence(C,q)$, measures how strongly chunks
belonging to course $C$ appear in the retrieved ranking. It gives more weight to
chunks appearing at higher ranks:

\[
RankedEvidence(C,q) =
\frac{\sum_{c \in C \cap R_q} \frac{1}{t_q + rank(c)}}
{\sum_{i=1}^{200} \frac{1}{t_q + i}}.
\]

Here, $rank(c)$ is the position of chunk $c$ in the top-200 retrieval ranking,
where rank 1 denotes the most similar chunk. The parameter $t_q$ represents
the number of relevant concepts identified in the query and acts as a smoothing
factor: highly ranked chunks contribute more strongly, while lower-ranked chunks
receive progressively lower weights.

The third component, $LessonCoverage(C,q)$, measures whether the relevant
evidence for course $C$ is distributed across multiple lessons rather than being
concentrated in a single fragment. For each lesson $l$ of course $C$, we define
$rank(l)$ as the best rank among the retrieved chunks belonging to that lesson:

\[
rank(l) = \min_{c \in l \cap R_q} rank(c).
\]

If no chunk of lesson $l$ appears in $R_q$, its contribution is zero. The lesson
coverage component is then defined as:

\[
LessonCoverage(C,q) =
\frac{\sum_{l \in C} \delta(l,q)}
{|L_C|},
\]

where $L_C$ is the set of lessons of course $C$ and

\[
\delta(l,q) =
\begin{cases}
\frac{1}{t_q + rank(l)} & \text{if } l \text{ contains at least one chunk in } R_q,\\
0 & \text{otherwise.}
\end{cases}
\]

This component rewards courses in which relevant evidence is spread across
several lessons, while penalising courses that contain only isolated relevant
fragments. The division by $|L_C|$ normalises the score with respect to the
number of lessons in the course, reducing the bias toward longer courses.

The final recommendation score is obtained by multiplying the three components.
This design favours courses that simultaneously: (i) account for a substantial
share of the total retrieved similarity; (ii) contain highly ranked relevant chunks;
and (iii) distribute relevant evidence across the internal structure of the course.
The resulting scores are used to rank courses and return the top recommendations
to the user.

\subsection{User Interface}

\begin{figure}[h!]
    \centering
    \includegraphics[width=1\linewidth]{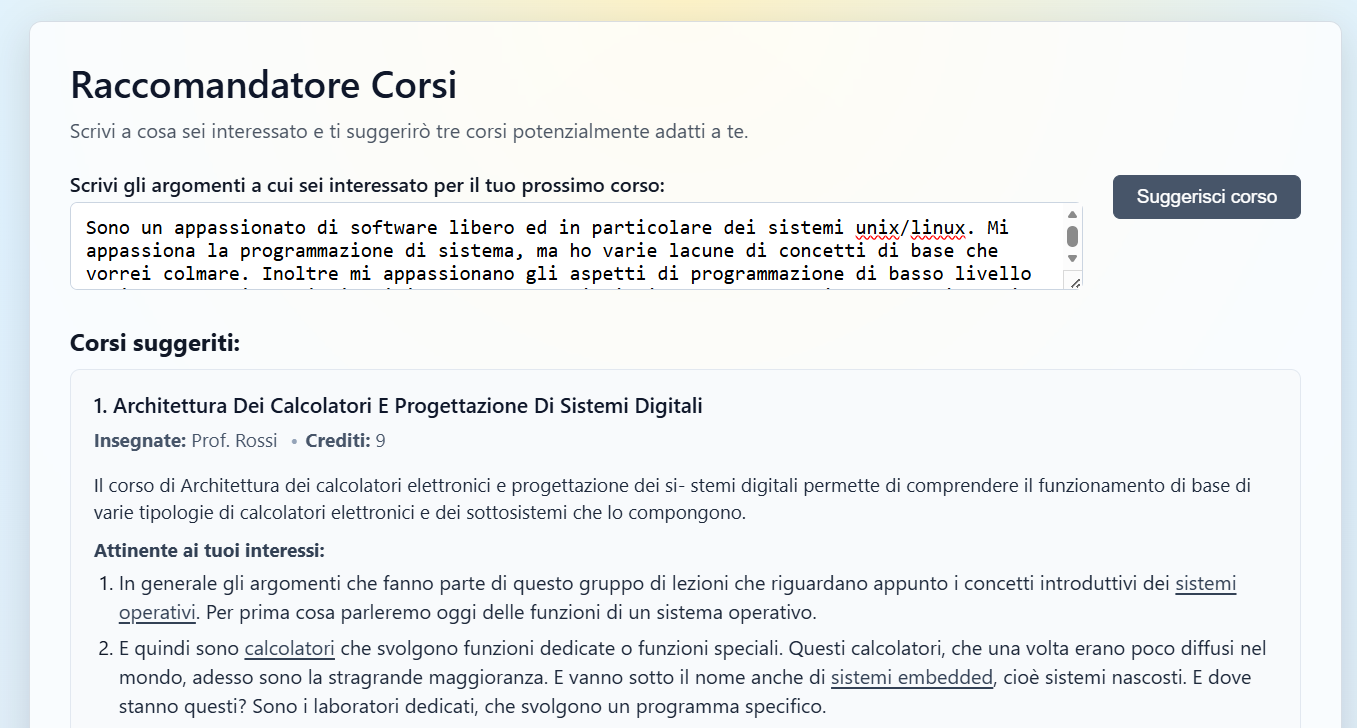}
    \caption{Example query and user interface output. The interface displays three recommended courses with metadata and supporting transcript chunks. Some information has been modified for copyright and privacy reasons.}
    \label{fig:ui}
\end{figure}

The system includes a lightweight web-based user interface designed to collect user queries and present course recommendations in an interpretable and structured manner. The interface consists of a single input field, where students enter a text description of their academic interests. Once submitted, the query is transmitted via an HTTP POST request to the other components. 
The interface then dynamically displays three recommended courses retrieved by the system,
each represented with its title, instructor, credits, course description, and
supporting transcript chunks. Figure~\ref{fig:ui} shows an example of
a student query and the corresponding ranked course recommendations. This presentation format allows students to inspect not only the thematic alignment of each course with their interests, but also the underlying semantic signals used by the model, thus supporting transparency and informed decision-making.

Overall, the interface functions as a thin interaction layer on top of the whole pipeline, enabling seamless communication between user inputs, retrieval processes, and output visualization while remaining intentionally minimal to avoid interfering with the underlying evaluation tasks.

\section{Evaluation}\label{sec:evaluation}

This section evaluates RAGEAR as a course-level recommender system. The evaluation is designed to assess two main aspects of the proposed approach: first, whether using full lecture transcripts provides more informative evidence than relying only on course-level metadata; second, whether the proposed RAGEAR aggregation function improves over a simpler transcript-based aggregation strategy. To this end, we compare three methods: a metadata-only dense retrieval baseline, a transcript-based normalized SumP aggregation baseline, and the full RAGEAR aggregation.

The evaluation is organised in three parts. Section~\ref{subsec:eval_protocol} describes the common evaluation protocol, including the compared methods, the relevance scale, and the ranking metrics. Section~\ref{subsec:human-llm-agreement} reports a human evaluation on a smaller balanced sample and measures the agreement between human and LLM-based relevance judgments. Finally, Section~\ref{subsec:large-scale-llm-evaluation} applies the validated LLM-based assessment protocol to the full set of 152 queries and reports the large-scale comparison of the three recommendation methods.

\subsection{Evaluation Protocol}\label{subsec:eval_protocol}


We evaluate RAGEAR as a course-level ranking system. Given a natural-language query expressing a student’s information need, each recommender returns an ordered list of courses. The evaluation compares three ranking strategies and isolates two contributions of the proposed approach: the use of lecture transcripts instead of course-level metadata, and the use of the full RAGEAR aggregation function instead of a simpler transcript-based aggregation.

We compare the following methods.

\textbf{Metadata baseline.} This baseline represents each course using only course-level information, namely the title, abstract, and available metadata. Dense retrieval is used to compute the semantic similarity between the student query and the course representation, and courses are ranked according to this score.

\textbf{Transcript Normalized SumP.} This method uses the full lecture transcripts associated with each course. Transcripts are segmented into chunks, and dense retrieval is applied at chunk level. Course-level relevance is computed through a normalized SumP aggregation: the score of a course is the sum of the similarity scores of its retrieved chunks, divided by the total similarity score of all retrieved chunks for the query. The underlying idea is related to classical evidence-combination methods in information retrieval, where multiple retrieval signals associated with the same item are combined by summing their scores, as in CombSUM \cite{fox1994combination}. Similar SumP-style strategies are still widely used in long-document retrieval, where document-level relevance is estimated by aggregating passage-level scores, for example in FirstP, MaxP, and SumP approaches \cite{Zhang2021150,Dai2019985}. Unlike the full RAGEAR aggregation, this baseline does not account for the rank positions of the retrieved chunks or for the distribution of relevant evidence across lessons.

\textbf{RAGEAR aggregation.} This is the full course-level aggregation method proposed in this paper. As in the previous method, dense retrieval is applied to transcript chunks. However, chunk-level scores are propagated to the course level using the complete RAGEAR scoring function, which combines three factors: (i) the proportion of the total retrieved similarity score associated with the course; (ii) the strength of the course evidence in the retrieved ranking, giving more weight to highly ranked chunks; and (iii) the distribution of relevant evidence across the lessons of the course. This rewards courses whose relevance is supported by multiple highly ranked chunks distributed across the course structure, rather than by isolated fragments.

All transcript-based methods use the same dense retrieval scores over the same transcript chunks. Therefore, the comparison between Transcript Normalized SumP and RAGEAR aggregation isolates the effect of the proposed course-level aggregation strategy.

The evaluation is conducted on student-like queries describing academic interests, learning goals, or orientation needs. The query set includes both metadata-friendly requests, which can often be answered from course titles or descriptions, and more fine-grained requests referring to concepts, skills, or topics that may only emerge from lecture content. 
Although the catalogue contains 34 courses, the task is non-trivial because many queries target fine-grained topics that may appear only in specific lessons rather than in course-level metadata.
For each query, each method returns a ranked list of recommended courses. For each candidate course, evaluators are shown the course title and an LLM-generated summary of the lecture transcripts, obtained through an API that processes the full set of lecture transcripts for that course.

Evaluators follow a two-step rating protocol, using the same rubric in both the human and LLM-based evaluations. First, they answer the binary question \textit{``Do you find this course relevant?''}. If the answer is \textit{no}, the course receives a relevance score of 0. If the answer is \textit{yes}, the evaluator assigns a score from 1 to 5, indicating increasing degrees of alignment with the query: 1 denotes a marginal relation, 2 partial relevance, 3 an acceptable recommendation, 4 a highly relevant recommendation, and 5 an ideal or near-ideal recommendation. In both settings, the evaluator is given the student query and the information needed to judge the relevance of a candidate course, but not the identity of the recommender that produced it.


We report standard ranking metrics at different cut-off levels. Since the system is intended to present a short list of recommendations, we focus primarily on cut-offs 1, 3, and 5. We compute Mean Reciprocal Rank (MRR), Precision@k, 
Mean Average Precision@k (MAP@k), and normalized Discounted Cumulative Gain (nDCG@k). For Precision 
and MAP, graded scores are binarized using a threshold: a course is considered relevant if its score is at least 3. For nDCG, we use the original graded relevance scores. MAP@k is truncated with normalization by the total number of relevant courses.

This protocol supports two research questions: whether lecture transcripts improve recommendation over metadata-only representations, and whether the full RAGEAR aggregation improves over normalized SumP aggregation of transcript-chunk scores. 
To support reproducibility, the query set, evaluation protocol, and aggregated results are made available in a public GitHub repository\footnote{\url{https://github.com/fpoggi/RAGEAR}}. Lecture transcripts are not released due to copyright and institutional constraints.

\subsection{
Human Evaluation and LLM Agreement}\label{subsec:humanvsllm}

\label{subsec:human-llm-agreement}

Before applying the LLM-based evaluation protocol to the full query set, we validate it on a smaller sample for which human relevance judgments are available. The goal of this step is not to replace human evaluation, but to assess whether LLM-based relevance judgments are sufficiently aligned with human judgments to support a larger-scale proxy evaluation of the ranking methods.

The human evaluation was designed as a balanced annotation task. We selected 20 queries and considered the ranked outputs produced by the three recommenders, resulting in 60 unique query--recommender output pairs. Each query--recommender output pair consists of one student query and the ranked list of courses returned by one of the three recommenders. Each participant evaluated six pairs: two produced by the metadata baseline, two by the Transcript Normalized SumP method, and two by the full RAGEAR aggregation. With 20 participants, this design yields 120 total human assessments. Since the 60 query--recommender pairs are distributed uniformly across participants, each pair is evaluated by two independent human annotators.

The same query--recommender output pairs were then assessed by the LLM judge, allowing us to measure agreement between human and LLM relevance judgments. All LLM-based assessments reported in this paper were conducted using GPT-4.1 nano\footnote{\url{https://developers.openai.com/api/docs/models/gpt-4.1-nano}}. 


We measure agreement using Kendall's $\tau$, Spearman's $\rho$, normalized Rank-Biased Overlap (RBO, $p=0.9$), and Jaccard similarity over the sets of relevant courses, using relevance score $\geq 3$ as threshold.

\begin{table}[t]
\centering
\caption{Metrics evaluating the statistical consistency of humans’ and LLM rating.}
\label{tab:human-llm-agreement}
\begin{tabular}{lcc}
\hline
\textbf{Metric} & \textbf{Mean} & \textbf{Std. Dev.} \\
\hline
Kendall's $\tau$ & \textbf{0.8} & \textbf{0.3} \\
Spearman's $\rho$ & \textbf{0.8} & \textbf{0.2} \\
RBO & \textbf{0.96} & \textbf{0.08} \\
Jaccard & \textbf{1.00} & \textbf{0.00} \\
\hline
\end{tabular}
\end{table}



Table~\ref{tab:human-llm-agreement} shows strong agreement between human and LLM assessments. Kendall's $\tau$ and Spearman's $\rho$ indicate high rank-level and score-level correlation, Jaccard reaches 1.0, and normalized RBO is approximately 0.96. This suggests that disagreements are mostly limited to minor reorderings of the same relevant courses.


Based on this agreement, we use the same LLM-based assessment protocol to scale the evaluation to the full set of 152 queries. We interpret the resulting scores as a controlled proxy for comparing ranking strategies, rather than as a substitute for a large-scale user study with students or academic advisors.


\subsection{Large-scale LLM-based Evaluation}
\label{subsec:large-scale-llm-evaluation}

After validating the agreement between human and LLM relevance assessments on the smaller human evaluation sample, we apply the same LLM-based judging protocol to the full query set. This large-scale evaluation includes 152 student-like queries, each representing an academic interest, a learning goal, or an orientation need. For each query, the three recommendation methods described in Section~\ref{subsec:eval_protocol} return a ranked list of candidate courses. The LLM judge then assigns a relevance score from 0 to 5 to each query--course pair, following the same rubric used in the human evaluation.
The goal of this experiment is to compare the ranking quality of the three methods under a common relevance assessment protocol. In particular, we evaluate whether lecture transcripts provide more informative evidence than course-level metadata, and whether the full RAGEAR aggregation improves over a simpler normalized SumP aggregation of transcript-chunk scores.




Table~\ref{tab:large-scale-results} reports the results. Overall, the transcript-based methods outperform the metadata-only baseline across all reported metrics. The \textit{Transcript Normalized SumP} baseline already improves over metadata-only retrieval, showing that full lecture transcripts contain useful fine-grained information that is not fully captured by course titles, abstracts, and metadata. For example, Precision@1 increases from 0.862 to 0.928, while nDCG@5 improves from 0.755 to 0.776.
The full RAGEAR aggregation further improves the results. Compared with the metadata baseline, RAGEAR obtains higher scores on all metrics, with particularly strong gains in the top-ranked results: Precision@1 increases by 10.69\%, nDCG@1 by 7.21\%, and MAP@5 by 17.09\%. This indicates that RAGEAR is more effective at placing highly relevant courses at the top of the recommendation list.

\begin{table}[ht]
\centering
\small
\setlength{\tabcolsep}{10pt}
\begin{tabular}{l|c|c|c}
\hline
\textbf{METRIC} & 
\textbf{METADATA} & 
\makecell{\textbf{TRANSCRIPT}\\\textbf{NORMALIZED}\\\textbf{SUMP ($\Delta$)}} & 
\textbf{RAGEAR ($\Delta$)}\\
\hline
MRR & 0.923 & 0.955 (+3.44\%) & \textbf{0.968 (+4.89\%)} \\
\hline
Precision@1 & 0.862 & 0.928 (+7.63\%) & \textbf{0.954 (+10.69\%)} \\
Precision@3 & 0.750 & 0.805 (+7.31\%) & \textbf{0.816 (+8.77\%)} \\
Precision@5 & 0.676 & 0.722 (+6.81\%) & \textbf{0.730 (+7.98\%)} \\
\hline
nDCG@1 & 0.773 & 0.795 (+2.85\%) & \textbf{0.828 (+7.21\%)} \\
nDCG@3 & 0.744 & 0.770 (+3.47\%) & \textbf{0.790 (+6.17\%)} \\
nDCG@5 & 0.755 & 0.776 (+2.80\%) & \textbf{0.795 (+5.29\%)} \\
\hline
MAP@3 & 0.239 & 0.267 (+11.70\%) & \textbf{0.280 (+16.82\%)} \\
MAP@5 & 0.326 & 0.366 (+12.13\%) & \textbf{0.382 (+17.09\%)} \\
\hline
\end{tabular}
\caption{Large-scale LLM-based evaluation on 152 queries. Percentages in parentheses report the relative improvement over the baseline. Bold values indicate the best-performing method for each metric.}
\label{tab:large-scale-results}
\end{table}

\vspace{-1cm}

The comparison between RAGEAR and \textit{Transcript Normalized SumP} isolates the contribution of the proposed aggregation function. Since both methods use the same dense retrieval scores over the same transcript chunks, their difference lies only in how chunk-level evidence is propagated to the course level. RAGEAR improves over the normalized SumP baseline on almost all reported metrics, especially those focused on the first positions of the ranking. For instance, nDCG@1 increases from 0.795 to 0.828, and MRR increases from 0.955 to 0.968. This suggests that considering not only the total amount of retrieved evidence, but also the rank of the relevant chunks and their distribution across lessons, helps produce better course-level rankings.

Since the recommender is intended to present only a short list of courses to the student, the most relevant metrics are those computed at cut-offs 1, 3, and 5. In this setting, RAGEAR consistently achieves the best performance. The gains are especially clear in MRR, Precision@1, MAP@5, and nDCG@1, suggesting that the proposed aggregation strategy is particularly effective in improving the quality of the top-ranked recommendations.

These results support two main conclusions. First, lecture-level content provides a stronger basis for course recommendation than course-level metadata alone. Second, aggregating transcript evidence through the full RAGEAR scoring function is more effective than relying only on the normalized sum of chunk-level similarity scores. The improvement is not due to a different retrieval model, since both transcript-based methods use the same dense retrieval scores, but to the way in which chunk-level evidence is combined into a course-level recommendation.

The evaluation should be interpreted as an offline comparison of ranking strategies. The LLM-based assessment enables a larger-scale evaluation than the human study, but it does not replace a longitudinal user study with students or academic advisors. Moreover, the experiments focus on the ranking component of RAGEAR, while symbolic filtering over credits, study plans, academic disciplines, and prerequisites is deterministic once the relevant information is encoded in the Knowledge Graph. Future evaluations should therefore assess the system in broader degree programmes and measure perceived usefulness, trust, and decision support in real course-selection scenarios.

\section{Conclusions and Future Work}\label{sec:conclusions}

This paper presented RAGEAR, a hybrid neural and symbolic recommender system for academic course recommendation. RAGEAR combines dense retrieval over full lecture transcripts with an ontology-based Knowledge Graph that models courses, lessons, transcript chunks, credits, study plans, academic disciplines, prerequisites, and other curricular information.

The main contribution of the system is a graph-aware aggregation function that propagates chunk-level retrieval evidence to course-level recommendations by considering score share, ranking strength, and distribution across lessons. Experimental results show that using lecture transcripts improves recommendation quality over metadata-only retrieval, and that the full RAGEAR aggregation further improves over a normalized SumP aggregation baseline, especially for top-ranked recommendations.

Future work will focus on extending the evaluation with larger human studies, applying the approach to broader academic programmes, integrating the prototype with operational student-facing services, and exploiting the semantic enrichment layer of the Knowledge Graph for explanation, concept-centric exploration, and more advanced forms of academic guidance.

\begin{credits}
\subsubsection{\ackname} 
We acknowledge financial support from the PNRR project Learning for All (L4ALL) funded by the Italian MIMIT (number: F/310072/01-05/X56).
\subsubsection{\discintname}
\end{credits}

%
%
%
\bibliographystyle{splncs04}
\bibliography{sn-bibliography}

@inproceedings{carroll2005named,
  title={Named graphs},
  author={Carroll, Jeremy J and Bizer, Christian and Hayes, Patrick and Stickler, Patrick},
  booktitle={Journal of Web Semantics},
  year={2005},
  doi = {10.1016/j.websem.2005.09.001}
}

@inproceedings{radford2023robust,
  title={Robust speech recognition via large-scale weak supervision},
  author={Radford, Alec and Kim, Jong Wook and Xu, Tao and Brockman, Greg and McLeavey, Christine and Sutskever, Ilya},
  booktitle={International conference on machine learning},
  pages={28492--28518},
  year={2023},
  organization={PMLR},
  doi = {10.5555/3618408.3619590}
}

@manual{explosion_spacy_2023,
  title        = {spaCy: Industrial-Strength Natural Language Processing in Python},
  author       = {{Explosion AI}},
  year         = {2023},
  note         = {Version 3.7.2},
  url          = {https://spacy.io}
}

@inproceedings{malhotra2022,
  title={Course recommendation using domain-based cluster knowledge and matrix factorization},
  author={Malhotra, Ishita and Chandra, Projit and Lavanya, R},
  booktitle={2022 9th International Conference on Computing for Sustainable Global Development (INDIACom)},
  pages={12--18},
  year={2022},
  organization={IEEE},
  doi = {10.23919/INDIACom54597.2022.9763281}
}

@article{esteban2020,
title = {Helping university students to choose elective courses by using a hybrid multi-criteria recommendation system with genetic optimization},
journal = {Knowledge-Based Systems},
volume = {194},
pages = {105385},
year = {2020},
issn = {0950-7051},
doi = {10.1016/j.knosys.2019.105385},
author = {A. Esteban and A. Zafra and C. Romero}
}

@ARTICLE{fernandez2020,
  author={Fernández-García, Antonio Jesús and Rodríguez-Echeverría, Roberto and Preciado, Juan Carlos and Manzano, José María Conejero and Sánchez-Figueroa, Fernando},
  journal={IEEE Access}, 
  title={Creating a Recommender System to Support Higher Education Students in the Subject Enrollment Decision}, 
  year={2020},
  volume={8},
  number={},
  pages={189069-189088},
  doi={10.1109/ACCESS.2020.3031572}}

@INPROCEEDINGS{sankhe2020,
  author={Sankhe, Viddhesh and Shah, Janice and Paranjape, Tejas and Shankarmani, Radha},
  booktitle={2020 IEEE International Conference on Computing, Power and Communication Technologies (GUCON)}, 
  title={Skill Based Course Recommendation System}, 
  year={2020},
  volume={},
  number={},
  pages={573-576},
  doi={10.1109/GUCON48875.2020.9231074}}

@inproceedings{obeid2018,
  title={An ontology-based recommender system for higher education},
  author={Obeid, Nizar and Lahoud, Chahine and El Khoury, Rami},
  booktitle={Proceedings of the International Conference on Advanced Learning Technologies},
  pages={45--49},
  year={2018},
  organization={IEEE},
  doi = {10.1145/3184558.3191533}
}

@ARTICLE{ibrahim2019,
  author={Ibrahim, Mohammed E. and Yang, Yanyan and Ndzi, David L. and Yang, Guangguang and Al-Maliki, Murtadha},
  journal={IEEE Access}, 
  title={Ontology-Based Personalized Course Recommendation Framework}, 
  year={2019},
  volume={7},
  number={},
  pages={5180-5199},
  doi={10.1109/ACCESS.2018.2889635}}

@inproceedings{li2024,
  title={CourseKG: A knowledge-graph approach for personalised course and learning-path recommendation},
  author={Li, Zheng and Wang, Mingyu and Zhao, Lin},
  booktitle={Proceedings of the International Conference on Educational Data Mining},
  pages={512--524},
  year={2024},
  organization={EDM},
  doi = {10.3390/app14072710}
}

@article{algarni2023systematic,
  title={Systematic review of recommendation systems for course selection},
  author={Algarni, Shrooq and Sheldon, Frederick},
  journal={Machine Learning and Knowledge Extraction},
  volume={5},
  number={2},
  pages={560--596},
  year={2023},
  publisher={MDPI},
  doi = {10.3390/make5020033}
}

@INPROCEEDINGS{hucontext,
  author={Hu, Lantao and Du, Zhao and Tong, Qiuli and Liu, Yongqi},
  booktitle={2013 IEEE 13th International Conference on Advanced Learning Technologies}, 
  title={Context-Aware Recommendation of Learning Resources Using Rules Engine}, 
  year={2013},
  volume={},
  number={},
  pages={181-183},
  doi={10.1109/ICALT.2013.56}}

@InProceedings{thuknow,
author="Nguyen, Thu Tran Minh
and Tran, Thinh Pham Quoc",
editor="Nguyen, Ngoc Thanh
and Iliadis, Lazaros
and Maglogiannis, Ilias
and Trawi{\'{n}}ski, Bogdan",
title="A Knowledge Graph Embedding Based Approach for Learning Path Recommendation for Career Goals",
booktitle="Computational Collective Intelligence",
year="2021",
publisher="Springer International Publishing",
address="Cham",
pages="66--78",
isbn="978-3-030-88081-1",
doi = "10.1007/978-3-030-88081-1_6"
}

@article{davisedu,
title = {Education in the era of Neurosymbolic AI},
journal = {Journal of Web Semantics},
volume = {85},
pages = {100857},
year = {2025},
issn = {1570-8268},
doi = {10.1016/j.websem.2024.100857},
author = {Chris Davis Jaldi and Eleni Ilkou and Noah Schroeder and Cogan Shimizu},
}

@inproceedings{Falco2014320,
	author = {Falco, Riccardo and Gangemi, Aldo and Peroni, Silvio and Shotton, David and Vitali, Fabio},
	title = {Modelling OWL ontologies with Graffoo},
	year = {2014},
	booktitle= {Lecture Notes in Computer Science (including subseries Lecture Notes in Artificial Intelligence and Lecture Notes in Bioinformatics)},
	volume = {8798},
	pages = {320 – 325},
	doi = {10.1007/978-3-319-11955-7_42},
	type = {Conference paper},
	source = {Scopus}
}

@incollection{Blomqvist2016,
  author = {Blomqvist, Eva and Hammar, Karl and Presutti, Valentina},
  booktitle = {Ontology Engineering with Ontology Design Patterns},
  editor = {Hitzler, Pascal and Gangemi, Aldo and Janowicz, Krzysztof and Krisnadhi, Adila and Presutti, Valentina},
  isbn = {978-1-61499-676-7},
  pages = {23--50},
  publisher = {IOS Press},
  series = {Studies on the Semantic Web},
  title = {{Engineering Ontologies with Patterns - The eXtreme Design Methodology}},
  volume = 25,
  year = 2016,
  doi = {10.3233/978-1-61499-676-7-23}
}

@inproceedings{fox1994combination,
  title={Combination of Multiple Searches},
  author={Fox, Edward A. and Shaw, Joseph A.},
  booktitle={The Second Text REtrieval Conference (TREC-2)},
  pages={243--252},
  year={1994},
  publisher={NIST}
}

@CONFERENCE{Dai2019985,
	author = {Dai, Zhuyun and Callan, Jamie},
	title = {Deeper text understanding for IR with contextual neural language modeling},
	year = {2019},
	journal = {SIGIR 2019 - Proceedings of the 42nd International ACM SIGIR Conference on Research and Development in Information Retrieval},
    pages = {985 – 988},
	doi = {10.1145/3331184.3331303},
	type = {Conference paper},
	source = {Scopus}
}

@ARTICLE{Zhang2021150,
	author = {Zhang, Xinyu and Yates, Andrew and Lin, Jimmy},
	title = {Comparing Score Aggregation Approaches for Document Retrieval with Pretrained Transformers},
	year = {2021},
	journal = {Lecture Notes in Computer Science},
	volume = {12657 LNCS},
	pages = {150 – 163},
	doi = {10.1007/978-3-030-72240-1_11},
	type = {Conference paper},
	source = {Scopus}
}

@article{wang2024multilingual,
  title={Multilingual E5 Text Embeddings: A Technical Report},
  author={Wang, Liang and Yang, Nan and Huang, Xiaolong and Yang, Linjun and Majumder, Rangan and Wei, Furu},
  journal={arXiv preprint arXiv:2402.05672},
  year={2024},
  doi={10.48550/arXiv.2402.05672}
}

@ARTICLE{Gangemi2026,
	author = {Gangemi, Aldo and Graciotti, Arianna and Meloni, Antonello and Nuzzolese, Andrea Giovanni and Presutti, Valentina and Reforgiato Recupero, Diego and Russo, Alessandro},
	title = {Text2AMR2FRED, converting text into RDF/OWL knowledge graphs via abstract meaning representation},
	year = {2026},
	journal = {Knowledge and Information Systems},
	volume = {68},
	number = {1},
	doi = {10.1007/s10115-025-02631-y},
	type = {Article},
	publication_stage = {Final},
	source = {Scopus}
}
%




\end{document}